\documentclass[runningheads]{llncs}
\usepackage[T1]{fontenc}
\usepackage{graphicx}
\usepackage{xcolor}
\usepackage{listings, listings-noir}
\usepackage{amsmath}
\usepackage{tabularx}
\usepackage[only,llbracket,rrbracket]{stmaryrd}

\newcommand{\Pv}[0]{$\mathcal{P}$}
\newcommand{\V}[0]{$\mathcal{V}$}
\newcommand{\seq}[1]{\langle#1\rangle}
\newcommand{\sem}[1]{\llbracket#1\rrbracket}

\newcommand{\ffadd}[1]{\lstinline|(ff.add |#1\lstinline|)|}
\newcommand{\ffmul}[1]{\lstinline|(ff.mul |#1\lstinline|)|}
\newcommand{\ffbitsum}[1]{\lstinline|(ff.bitsum |#1\lstinline|)|}

\newcommand{\imul}[1]{\lstinline|(* |#1\lstinline|)|}

\newcommand{\smteq}[1]{\lstinline|(= |#1\lstinline|)|}
\newcommand{\smtnot}[1]{\lstinline|(not |#1\lstinline|)|}
\newcommand{\smtimp}[1]{\lstinline|(=> |#1\lstinline|)|}
\newcommand{\smtlt}[1]{\lstinline|(< |#1\lstinline|)|}

\newcommand{\assert}[1]{\lstinline|verify\_assert(|#1\lstinline|)|}

\newcommand{\nave}[0]{NAVe}

\begin{document}
\title{Formally Verifying Noir Zero Knowledge Programs with \nave{}}
%
%
\author{Pedro Antonino \and
Namrata Jain}
\authorrunning{P. Antonino and N. Jain}
%
\institute{The Blockhouse Technology Ltd., Oxford, UK \\
\email{\{pedro,namrata\}@tbtl.com}\\
\url{www.tbtl.com}}
\maketitle              
\begin{abstract}
Zero-Knowledge (ZK) proof systems are cryptographic protocols that can (with overwhelming probability) demonstrate that the pair $(X, W)$ is in a relation $R$ without revealing information about the private input $W$. This membership checking is captured by a complex arithmetic circuit: a set of polynomial equations over a finite field. ZK programming languages, like Noir, have been proposed to simplify the description of these circuits. A developer can write a Noir program using traditional high-level constructs that can be compiled into a lower-level ACIR (Abstract Circuit Intermediate Representation), which is essentially a high-level description of an arithmetic circuit. In this paper, we formalise some of the ACIR language using SMT-LIB and its extended theory of finite fields. We use this formalisation to create an open-source formal verifier for the Noir language using the SMT solver cvc5. Our verifier can be used to check whether Noir programs behave appropriately. For instance, it can be used to check whether a Noir program has been properly constrained, that is, the finite-field polynomial equations generated truly capture the intended relation. We evaluate our verifier over 4 distinct sets of Noir programs, demonstrating its practical applicability and identifying a hard-to-check constraint type that charts an improvement path for our verification framework.



\keywords{Zero Knowledge Proofs  \and Zero Knowledge Programs \and Noir \and Privacy \and Formal Verification}
\end{abstract}

\section{Introduction}

Zero Knowledge (ZK) proof systems were initially introduced as interactive protocols~\cite{Goldwasser85,goldwasser89,Goldreich91} where a \emph{prover} \Pv{} convinces a \emph{verifier} \V{} of a statement of the form $R(X,W)$ where $R$ is a relation (known to \Pv{} and \V{}), $X$ is an input common to both \Pv{} and \V{}, and $W$ is an input private to the prover (commonly referred to as a \emph{witness}); the protocol's execution must reveal nothing about $W$ apart from the fact that $R(X,W)$ holds. These proof systems have been made non-interactive \cite{Blum88,Groth10} and (their generated proof) succinct \cite{Gennaro12,Parno16,Groth16,Maller19}. 
ZK proof systems have gained renewed interest with the advent of blockchains and the growth of the Web3 ecosystem. They have been promoted as a technology to deliver privacy~\cite{aztec,starknet} and, even, improve the efficiency of blockchains~\cite{polygon,beam}.    


In these proof systems, the relation $R$ is typically captured by an arithmetic circuit: a set of constraints (i.e. polynomial equations) over a finite field that capture the relationship between $X$ and $W$. If one is not familiar with such systems, describing these circuits can be challenging. To simplify the use of these systems and increase their adoption, Domain-Specific Languages (DSLs) targeting arithmetic circuits have been developed. These \emph{ZK programming languages} provide high-level constructs and an infrastructure to compile these constructs into arithmetic circuits used by ZK proof systems; Noir~\cite{noir-lang}, Cairo~\cite{cairo-lang}, and (in a lower level of abstraction) Circom~\cite{circom-lang} are examples of such languages. Simply put, they can be used to create a high-level program $P$ that captures $R$: if $R(X,W)$ holds, the program $P$ successfully runs with parameters $(X,W)$. 

The Noir language~\cite{noir-lang} was proposed by the Aztec community~\cite{aztec}. It can be used to both create independent ZK programs and private smart contracts for the Aztec network; smart contracts are a type of program that manages digital assets sometimes worth large amounts of money. Noir programs are compiled into an intermediate language called ACIR (Abstract Circuit Intermediate Representation), which is then used by a ZK proof system. This intermediate representation has opcodes capturing finite-field polynomial equations, but it also has higher-level opcodes, such as \emph{range check} --- a constraint that ensures that a finite-field value can be represented with a determined number of bits --- and \emph{brillig call} --- an opcode that allows the execution of \emph{unconstrained code}. Unconstrained code is a feature of these languages designed to help the prover produce a proof, but it does not contribute to the underlying constraints capturing the relation $R$. The programmer is expected to include separately the appropriate matching constraints using, for instance, the \lstinline{assert} command in Noir.

This new programming paradigm is prone to several subtle types of bugs~\cite{zk-bug-tracker,Chaliasos24}. For instance, a lack of appropriate constraining in the Noir program, say if it executes unconstrained code that was not appropriately constrained, can lead to a divergence between the program $P$ and the underlying relation it is trying to capture $R$. This type of bug has already led to serious real-world vulnerabilities; see~\cite{tornado-cash}, for instance. ZK programs can also suffer from other types of bugs, such as non-determinism --- the underlying relation was meant to be a function, but it is not --- and even compiler bugs --- the constraints generated do not appropriately capture the behaviour of the high-level programming constructs. The subtlety of these bugs and the increased application of ZK programs in high-stakes contexts --- protecting users' private information or managing large sums of digital assets --- make them a perfect target for formal verification.

In this paper, we formalise a subset of ACIR using SMT-LIB~\cite{SMTLIB} and propose \nave: the Noir Acir (formal) Verifier; not to be confused with the underlying ZK proof system verifier. We encode the semantics of a subset of the ACIR opcodes into SMT-LIB. We propose two encodings: one based on the standard theory of integer arithmetic and another based on a non-standard theory of finite fields~\cite{Ozdemir23,Ozdemir24} that is implemented in the cvc5 solver~\cite{Barbosa22}. We use this formalisation(s) and the convention of capturing verification goals/proof obligations using the Noir function \lstinline{verify_assert(b: bool)} to create our open-source formal verifier \nave. Our verifier focuses on identifying unconstrained bugs: a Noir developer can use \lstinline{verify_assert} call at the program's exit point to specify the desired relation $R$ it is expected to capture. 
We have integrated our verifier into Nargo, Noir's package manager, to ease developers into using it. We have evaluated our verifier in some test programs that are part of the Noir language infrastructure to assess its efficiency in practice. We analyse its performance using both of our encodings with encouraging results. This evaluation suggests that: (a) some types of constraints, induced by opcodes like range checks, can significantly increase verification time and that (b) our encodings are complementary in the sense that some programs are more quickly analysed using one and some others using the other. So, our evaluation charts a path for future work: we need to propose abstractions/approximations to address (a) and some form of analysis of a Noir program with both encodings concurrently to benefit from (b).

There has been some work on the automated formal verification of ZK programs~\cite{Ozdemir22,Pailoor23,Stephens25}, none of which targets the Noir language and its particularities. Noir has high-level constructs with complex semantics. For instance, unlike other languages, Noir/ACIR supports high-level memory operations, unconstrained functions, and bit-level operations like range checks, bitwise operators, etc. Lampe~\cite{lampe} captures the semantics of Noir using Lean 4~\cite{Lean4}. Unlike our automated verifier, the Lampe user has to interactively drive the proof of program properties. We chose to formalise ACIR instead of Noir directly, as the former should be more stable and undergo fewer evolutionary changes than the latter. Also, our formalisations could be used by other languages built to target ACIR.

To sum up, our contributions are:
\begin{itemize}
    \item Formalisations of ACIR (and indirectly Noir) using SMT-LIB standard theory of integer arithmetic and the non-standard theory of finite fields~\cite{Ozdemir23}.
    \item An open-source formal verifier, \nave, for the Noir programming language based on our formalisation, which is integrated into the popular Noir package manager (and build tool) Nargo.
    \item An evaluation of \nave{} on Noir test programs that demonstrate its practical applicability and identifies a class of complex constraints and solver/encoding combinations, charting a path for future improvement.
\end{itemize}

\noindent
\textit{Outline.} In Section~\ref{sec:background}, we present the necessary background to make our paper self-contained. Section~\ref{sec:verifier} introduces our formalisation(s), verifier, and its evaluation. Section~\ref{sec:related} presents some related work and Section~\ref{sec:conclusion} our concluding remarks.

\section{Background}
\label{sec:background}

In this section, we introduce some concepts to make our paper self-contained.

\subsection{SMT-LIB, CVC5, and Finite Field Theory}

Satisfiability Modulo Theories (SMT) generalises boolean satisfiability to other theories such as integer arithmetic. An SMT formula involves predicates in these theories that can be combined using boolean operators. SMT solvers, like cvc5~\cite{Barbosa22}, are tools implementing decision procedures to verify the satisfiability of such formulas. SMT-LIB~\cite{SMTLIB} is a library that standardises some of these theories, providing the semantics and syntax for those. It has become the de-facto standard syntax to represent these formulas. So, even theories that are not SMT-LIB standard use its syntax. For instance, \lstinline{(declare-const x Int)} can be used to declare a constant \lstinline{x} of type/sort \lstinline{Int}, whereas \lstinline{(assert (or (= x 0) (= x 1)))} declares the constraint that \lstinline{x} is either 0 or 1. And one can use \lstinline{(+ x y)}, \lstinline{(* x y)}, and \lstinline{(mod x y)} to construct the corresponding typical integer operations.

The cvc5 solver also supports the non-standard theory of finite fields of prime order~\cite{cvc5-ff} described in~\cite{Ozdemir23}. A finite field of prime order $p$ is isomorphic to the integers modulo $p$. For integer $x$ and positive integer $y$, we use $(x \mod y)$ to denote the unique integer $r$ such that $q$ is an integer, $x = qy + r$ and $0 \leq r < y$. Let $p$ be a prime number, $F$ be the sort representing a field of prime order $p$, $N$ be an integer, $x_i$ be terms capturing field values in the same field $F$, and $[x]$ be the integer denoted by the finite-field term $x$.
\begin{itemize}
    \item \lstinline{(_ FiniteField} $p$\lstinline{)} denotes the field of prime order $p$ sort, i.e. the set of integers modulo $p$.
    \item \lstinline{(as ff}$N$ $F$\lstinline{)} denotes the finite field element $N$, i.e. the integer $(N \mod p)$.
    \item \lstinline{(ff.add} $x_1\ \ldots\ x_n$\lstinline{)} captures the n-ary field addition operation. This term denotes the integer $([x_1] + \ldots + [x_n] \mod p)$.
    \item \lstinline{(ff.mul} $x_1\ \ldots\ x_n$\lstinline{)} captures the n-ary field multiplication operation. This term denotes the integer $([x_1] \times \ldots \times [x_n] \mod p)$.
    \item \lstinline{(ff.bitsum} $x_1\ \ldots\ x_n$\lstinline{)} captures the n-ary bitsum operation. This term denotes the integer $(2^0[x_1] + \ldots + 2^{n-1}[x_n] \mod p)$.
    \item \lstinline{(= }$x$ $y$\lstinline{)} denotes the equality operator. For the integers denoted by $[x]$ and $[y]$, this equality holds if and only if $[x] = [y]$.
\end{itemize}

The isomorphism between the field of prime order $p$ and the integers modulo $p$ means that one can encode finite-field terms as non-linear integer arithmetic terms and use a solver for this theory to check for satisfiability; this problem is undecidable in general, however, whereas it is decidable for finite-field terms.

\subsection{Noir}

Noir~\cite{noir-lang} is a ZK programming language that was proposed by the Aztec~\cite{aztec} community. It can be used to either construct stand-alone programs or smart contracts for the Aztec network. Noir is a high-level programming language whose syntax is inspired by Rust~\cite{rust-lang}. Like in Rust, abstract data types can be created via \emph{structs} and their corresponding implementation. Unlike traditional programming languages, Noir has a built-in \lstinline{Field} type denoting a finite field and field operations like addition and multiplication. The actual field depends on the (backend) proof system used. For Noir's default backend, a field element would be a 254-bit integer that spans the Grumpkin curve~\cite{grumpkin}\footnote{The finite field has order $218882428718392752222464057452572750885483644004160\allowbreak34343698204186575808495617$.}. Another peculiarity of ZK programming languages is the distinction between constrained and unconstrained code. While the former gives rise to arithmetic-circuit constraints, the latter does not, and it has to be combined with separate constraints manually added by the developer. Unconstrained code can greatly reduce the size and complexity of the underlying arithmetic circuit generated, improving proving and verification times. 

\begin{figure}[!t]
    \centering
    \begin{minipage}[t]{0.45\textwidth}
\begin{lstlisting}[mathescape]
fn main(x: Field) -> pub Field {
    let invx: Field = inv(x);
    invx
}

fn inv(x: Field) -> Field {
    let mut cur = x;
    let mut y = 1;
    let rep: [u1; 253] = 
        [1,1,$\ldots omitted\ldots$];
    for i in 0..253 {
        if (rep[i] == 0) {
            cur = cur * cur;       
        } else {
            y = y * cur;
            cur = cur * cur;   
        }
    }
    y * cur
}
\end{lstlisting}    
\end{minipage}\hfill%
\begin{minipage}[t]{0.45\textwidth}
\begin{lstlisting}
fn main(x: Field) -> pub Field {
    let mut invx = 0;
    unsafe {
        invx = invu(x);
    }
    assert(invx * x == 1);
    invx
}

unconstrained fn invu(x: Field) 
    -> Field {
        inv(x)
}
\end{lstlisting}
\end{minipage}
    \caption{Noir programs: constrained vs unconstrained.}
    \label{fig:noir_programs}
\end{figure}

We briefly present the Noir language and the advantages of unconstrained code with the Noir programs in Figure~\ref{fig:noir_programs}. The multiplicative inverse of a non-zero element $x$ in a finite field can be calculated by computing $x^{p-2}$, by Fermat's little theorem, where $p$ is the prime order of the field. Both the left- and right-hand side programs, denoted by the respective \lstinline{main} functions, compute the public value \lstinline{x}$^{p-2}$, denoted by the \lstinline{pub Field} return type, from the private value \lstinline{x}. With this program, a prover can convince a verifier that it knows \lstinline{x} for the public value \lstinline{x}$^{p-2}$ without revealing \lstinline{x}.\footnote{These programs can be found at \url{https://github.com/pedrotbtl/nave}. They are a didactic tool to demonstrate the discrepancy between constrained and unconstrained code. Noir has built-in unconstrained code and constraints to calculate $1/x$, and a verifier could compute \lstinline{x} from \lstinline{x}$^{p-2}$, bypassing the program's privacy guarantees.} The left-hand side Noir program uses constrained code to carry out this computation, whereas the right-hand side one uses unconstrained code. In Noir, the \lstinline{let} keyword is used to declare a variable, and the \lstinline{mut} modifier denotes that the variable is mutable. The \lstinline{inv} function uses iterative squaring~\cite{Knuth97} to compute $x^{p-2}$: a \lstinline{for}-loop iterates over the binary iterative decomposition \lstinline{rep} of $p-2$ to carry out this exponentiation. For this example, we use $p$, the prime order of the Grumpkin base field, and we omit this 253-bit array from the code for brevity. The right-hand side uses the same \lstinline{inv} (inside function \lstinline{invu}) code but in an unconstrained way and relies on the \lstinline{assert(invx * x == 1)} statement to ensure the output \lstinline{invx}, calculated in an unconstrained way, is the inverse of \lstinline{x}; the \lstinline{assert} command generates constraints for the underlying arithmetic circuit. While the left-hand side Noir program generates 379 constraints, the right-hand side one only generates 2. In ZK programs, unconstrained code helps the prover in generating a proof. In this case, in a ZK proof system, the prover needs to compute the inverse of \lstinline{x} to create a proof. For the ZK verification, however, the verifier only needs to check that \lstinline{invx} is the inverse of \lstinline{x}.

\section{\nave: a formal verifier for Noir programs}
\label{sec:verifier}

In this section, we formalise Noir's ACIR, propose a formal program verifier called \nave{}, and conduct an evaluation to assess \nave's practical applicability. 

\subsection{ACIR formalisation}

A Noir program is compiled into an intermediate \emph{ACIR program}, which is later translated into an arithmetic circuit targeting a specific ZK proof system. ACIR is designed to decouple the frontend high-level ZK language from the backend ZK proving system. It was created by the Aztec community for Noir, but it is designed to be high-level-language agnostic. We decided to target ACIR as opposed to Noir directly to: (a) benefit from the infrastructure used to compile Noir programs into ACIR and the lower-level nature of ACIR makes it simpler to capture its semantics, (b) for as long as this intermediate language is stable, and it should be stabler than Noir, our formal verifier should remain compatible with Noir as it evolves into new versions, and (c) as ACIR was designed to be frontend agnostic, our ACIR semantics could be used in the future to verify other high-level ZK languages targeting ACIR as their intermediate representation.

The ACIR language has 6 opcodes: \lstinline{AssertZero}, \lstinline{BlackBoxFuncCall}, \lstinline{MemoryOp}, \lstinline{MemoryInit}, \lstinline{BrilligCall}, and \lstinline{Call}. In this work, we only formalise a subset of ACIR. We do not formalise the \lstinline{Call} opcode and some of the behaviour of the \lstinline{BlackBoxFuncCall} opcode; we leave that as future work. Functions in Noir are typically inlined, so the \lstinline{Call} opcode is only used in some specific, and somewhat intricate, proving scenarios using recursive proving. As for the \lstinline{BlackBoxFuncCall}, some of its behaviour is quite complex as we detailed later.


For a given finite field, an ACIR program consists of a list of ACIR opcodes $\seq{op_1,\ldots,op_m}$ operating on field variables, which are called \emph{witnesses}, and a list of brilling functions; the latter is not relevant for this paper. Opcodes can be seen as creating constraints limiting which values the witnesses can take, and a valid assignment for the witnesses captures an execution of the ACIR program. In the following, we formalise the semantics of opcodes via SMT-LIB, namely, we define the constraints that capture each of ACIR opcodes' behaviour. We propose a translation of ACIR opcodes both into a non-standard SMT-LIB theory of finite fields (see Section~\ref{sec:background}) and the standard theory of integers.

Some opcodes operate over an ACIR \lstinline{Expression}: a datatype representing a polynomial of degree at most 2. Let $r_i,s_i,q$ be (literal) field elements and $a_i,b_i,c_i$ witnesses. This datatype is a triple containing: an $n$-long list of triples of the form $(r_i,a_i,b_i)$, an $m$-long list of pairs $(s_i,c_i)$, and a field element $q$, and it represents the field polynomial $\sum^n_{i=1} r_ia_ib_i + \sum^m_{j=1} s_jc_j + q$. We capture this expression in SMT-LIB as follows. Both ACIR and SMT-LIB variables span over the same finite field $F$. We use $\sem{t}_f$ to denote our encoding for ACIR term $t$ (be it an opcode or expression) using the finite-field theory. To simplify our presentation, we omit the encoding of field elements: the value $v$ is, of course, encoded as \lstinline{(as ff}$v$ $F$\lstinline{)} where $F$ is a sort definition for the appropriate field.
\begin{flalign*}
&\sem{(\seq{(r_1, a_1, b_1),\ldots, (r_n, a_n, b_n)},\seq{(s_1, c_1),\ldots, (s_m, c_m)}, q) : \text{\lstinline{Expression}}}_{f} \equiv && \\ 
& \quad  \text{\lstinline{(ff.add} \ffmul{ $r_1$ $a_1$ $b_1$} $\ldots$ \ffmul{ $r_n$ $a_n$ $b_n$}} \\
& \quad \quad \quad  \text{\ffmul{ $s_1$ $c_1$} $\ldots$ \ffmul{ $s_m$ $c_m$} $q$\lstinline|)|}
\end{flalign*}

We rely on the isomorphism between modular integer arithmetic and field arithmetic to propose an alternative ACIR encoding via SMT-LIB standard theory of integers; we use $\sem{t}_i$ to denote this formalisation. In the following SMT-LIB formula, $r_i,s_i,q$ are (literal) integers and $a_i,b_i,c_i$ are integer variables.
\begin{flalign*}
&\sem{(\seq{(r_1, a_1, b_1),\ldots, (r_n, a_n, b_n)},\seq{(s_1, c_1),\ldots, (s_m, c_m)}, q) : \text{\lstinline{Expression}}}_{i} \equiv && \\ 
& \quad \text{\lstinline{(mod} \lstinline{(+} \imul{ $r_1$ $a_1$ $b_1$} $\ldots$ \imul{ $r_n$ $a_n$ $b_n$}\imul{ $s_1$ $c_1$} $\ldots$ \imul{ $s_m$ $c_m$} $q$\lstinline|)| $p$\lstinline|)|}
\end{flalign*}

The \lstinline{AssertZero} opcode gives rise to a polynomial-equation constraint. For an ACIR expression $exp$, this opcode induces the polynomial-equation constraint $\sum^n_{i=1} r_ia_ib_i + \sum^m_{j=1} s_jc_j + q = 0$. The integer translation is the same, but for the integer expression encoding.
\begin{flalign*}
&\sem{\text{\lstinline{AssertZero}}(exp)}_{f} \equiv \text{\lstinline{(=} $\sem{exp}_f$ 0\lstinline|)|} &&
\end{flalign*}

For instance, the opcode \lstinline{AssertZero}$(\seq{(1,w_1,w_2)},\seq{},-1)$ captures the constraint $w_1w_2 - 1 = 0$, namely, that $w2$ is the multiplicative inverse of $w_1$. This type of opcode is generated, for instance, by Noir statement \lstinline{assert(invx * x = 1)}.


The \lstinline{BlockBoxFuncCall} opcode allows the execution of \emph{black box functions}. They encompass behaviours that cannot usually be efficiently captured by polynomial constraints such as the ones generated by \lstinline{AssertZero} opcodes. They are designed to give the underlying ZK proof system some flexibility in providing its own optimised circuits to capture these functions. ACIR supports 14 block box functions: varying from range checking to complex hash functions and encryption computations. In this paper, we have only provided a formalisation of the $\textit{Range}$ black box function; we leave the formalisation of the other functions as future work. For a witness $x$ and an integer $n$, the $\textit{Range}(x,n)$ function checks that the value of $x$ can be encoded as an $n$-bit-long unsigned integer. We use a binary representation of $x$ to implement this check. We use new field variables $y_1,\dots,y_n$ to encode a bitsum representation for $x$, that is, $x = \sum^n_{i=1} 2^{i-1}y_i$, and we enforce that $y_i$ represent a binary value with constraint $y_i(y_i - 1) = 0$.
\begin{flalign*}
&\sem{\text{\lstinline{BlockBoxFuncCall}}(\textit{Range}(x,n))}_{f} \equiv && \\ 
 &\quad \text{\smteq{ \ffbitsum{ $y_1$ $\ldots$ $y_n$} $x$}} \land \\
  &\quad \quad \text{\smteq{ \ffmul{ $y_1$ \ffadd{ $y_1$ $-1$}} 0} $\land \ldots \land$ \smteq{ \ffmul{ $y_n$ \ffadd{ $y_n$ $-1$}} 0}}
\end{flalign*}
Similar black box functions, like $\textit{AND}$ and \textit{XOR}, representing bit-level logical $and$ and $xor$, respectively, can be formalised using this binary representation.

Unlike the finite-field encoding, the integer-arithmetic formalisation can rely on the intrinsic order of the integers to implement this check.
\begin{flalign*}
&\sem{\text{\lstinline{BlockBoxFuncCall}}(\textit{Range}(x,n))}_{i} \equiv \text{\smtlt{ $x$ $2^n$}} &&
\end{flalign*}

The \lstinline{MemoryOp} and \lstinline{MemoryInit} opcodes give ACIR the ability to create, access, and mutate \emph{arrays of (new) witnesses}. These arrays are identified by a \emph{memory block id}. Our encodings also need a way to track when arrays have been mutated. We use $m_{id,i,j}$ to represent the value of the element $i$ of array (i.e. memory block) $id$ after $j$ mutations. For instance, $m_{2,1,0}$ gives the element at index $1$ for array $2$ at initialisation, whereas $m_{2,1,1}$ gives the value for the same element and array after the first write. We translate/encode memory operations sequentially and keep track of the mutation index of arrays with the map $mi$. So, $mi$ is part of the translation context, and it is updated when an array is created and updated. We do not add it to the translation rules for the sake of conciseness.

The \lstinline{MemoryInit} opcode creates an array of new witnesses.\footnote{ACIR allows for different memory block types. Here, we only formalise the more general \emph{Memory} type.} Given a block identifier (an unsigned 32-bit integer) $id$ and witnesses $x_1,\ldots,x_n$ as inputs, for each input witness $x_i$, this opcode generates a new (fresh) witness $m_{id,i,0}$ with the same value as $x_i$. As part of this operation, the mutation index map is updated as follows $mi' = mi \oplus \{id \rightarrow 0\}$, where $mi'$ and $mi$ are the value of the map after and before the translation, respectively, and $\oplus$ adds a new maplet to the map or updates it if there already existed another maplet with the same $id$.
\begin{flalign*}
&\sem{\text{\lstinline{MemoryInit}}(id, \seq{x_1,\ldots,x_n})}_{f} \equiv \text{\smteq{ $m_{id,1,0}$ $x_1$} $\land  \ldots \land$ \smteq{ $m_{id,n,0}$ $x_n$}}  &&
\end{flalign*}

The \lstinline{MemoryOp} opcode can be used to read from and write into memory blocks. Its inputs are a block identifier $id$, a memory operation $op$ (can be $0$ for a read or $1$ for a write), an index ACIR expression $ind$ and a value ACIR expression $val$. A read copies the value at index $ind$ into the witness given by $val$. Note how we use the value $mi(id)$ to identify the witnesses representing the current state of the array and how the implication identifies the correct index to be accessed. There is no change in the state of the array, and hence to $mi$.
\begin{flalign*}
&\sem{\text{\lstinline{MemoryOp}}(id, op=0, ind, val)}_{f} \equiv \text{\smtimp{ \smteq{ $\sem{ind}_f$ 1} \smteq{ $m_{id,1,mi(id)}$ $\sem{val}_f$}}} \land && \\
& \quad \ldots\land \text{\smtimp{ \smteq{ $\sem{ind}_f$ n} \smteq{ $m_{id,n,mi(id)}$ $\sem{val}_f$}}}
\end{flalign*}

A write mutates the value at index $ind$ to $val$. Note how we use the value $mi(id)$ and $mid(id)+1$ to identify the witnesses representing the current and the next state of the array, respectively. The first set of conjunctions is very similar to a read, apart from the fact that they constrain the next state of the array $mi(id)$, and they define what happens to the element being written to. The second set of conjuncts defines what happens to the elements of the array that are not written to; that is, they retain their previous value. The mutation mapping $mi$ is also updated: $mi' = mi \oplus \{id \rightarrow mi(id) + 1\}$.
\begin{flalign*}
&\sem{\text{\lstinline{MemoryOp}}(id, op=1, ind, val)}_{f} \equiv \text{\smtimp{ \smteq{ $\sem{ind}_f$ 1} \smteq{ $m_{id,1,mi(id)+1}$ $\sem{val}_f$}}} \land && \\
& \quad \ldots\land \text{\smtimp{ \smteq{ $\sem{ind}_f$ n} \smteq{ $m_{id,n,mi(id)+1}$ $\sem{val}_f$}}} \land && \\
& \quad \text{\smtimp{ \smtnot{ \smteq{ $\sem{ind}_f$ 1}} \smteq{ $m_{id,1,mi(id)+1}$ $m_{id,1,mi(id)}$}}} \land && \\
& \quad \ldots\land \text{\smtimp{ \smtnot{ \smteq{ $\sem{ind}_f$ n}} \smteq{ $m_{id,n,mi(id)+1}$ $m_{id,n,mi(id)}$}}}
\end{flalign*}

The modular integer formalisation for memory opcodes mirrors the above definitions by using corresponding integer witnesses and expression encoding.

Finally, \lstinline{BrilligCall} opcodes capture \emph{unconstrained code}. As we are interested in analysing and formalising constrained code, whose execution integrity is enforced by the underlying ZK proof system, we do not translate these opcodes, or perhaps more appropriately, they are translated as the always satisfied constraint \lstinline{true}.

\subsection{Verifier}

\begin{figure}[!b]
    \centering
    \includegraphics[width=\textwidth]{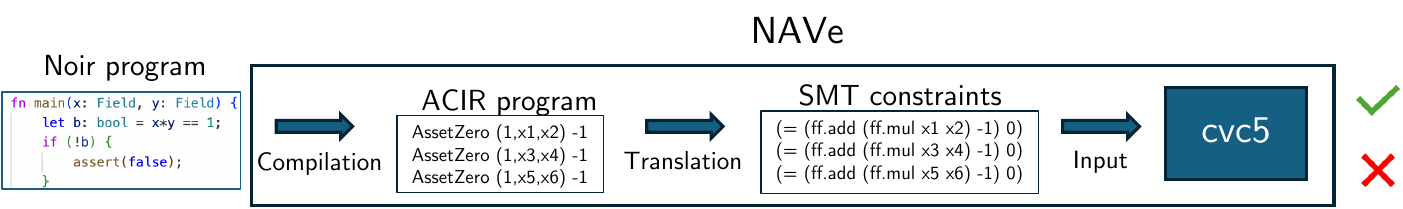}
    \caption{Overview of \nave{} formal verifier.}
    \label{fig:nave}
\end{figure}

We propose \nave: a formal verifier for Noir programs based on our SMT-LIB semantics --- its high-level architecture is given in Figure~\ref{fig:nave}. It relies on the Noir infrastructure for compiling Noir programs into ACIR, and cvc5~\cite{noir-lang} to solve the SMT constraints generated by our semantics (i.e. SMT-LIB translations). Our verifier is publicly available in our repository.\footnote{\url{https://github.com/pedrotbtl/nave}}

We use the basic Noir program in Figure~\ref{fig:basic_program} to demonstrate the usage of our formal verification framework. A Noir program can be annotated with a \assert{$cond$}. Unlike a built-in Noir \lstinline{assert} which creates a constraint, this \emph{formal verification function} creates a proof obligation: the condition $cond$ must hold at the point of that statement. The $cond$ is a \emph{side-effect-free} predicate over the variables in scope at that point in the program. For now, this is a simple, unquantified Noir expression. This verification function must be declared by the user of \nave{}, like in the basic example program. This approach minimises the changes to the Noir language infrastructure. We do not have to add new language constructs (and define their parsing, compilation, etc) to track verification conditions. Instead, we intercept at the ACIR-level a call to this specific function and do the appropriate translation into a verification condition. Moreover, this function does not modify the behaviour of the Noir program at all. As it is an unconstrained function, it does not contribute constraints to the underlying arithmetic circuit, and its execution does nothing. So, developers can annotate their programs with these formal verification constructs without impacting their corresponding ZK proof generation and verification.

\begin{figure}[!t]
    \centering
\begin{lstlisting}
fn main(x : Field) {
    assert(x * (x - 1) == 0);
    unsafe { verify_assert((x == 0) | (x == 1)); }
}

unconstrained fn verify_assert(b: bool) {}
\end{lstlisting}    
    \caption{Basic example program.}
    \label{fig:basic_program}
\end{figure}

\nave{} is implemented (in a public fork of and) as a command (\lstinline{formal-verify}) of the Nargo package manager. We hope that by using a tool that is familiar to Noir developers, they will be more comfortable making use of \nave{}. For a Noir program $P$, we use the Noir compiler to generate the ACIR program $A$. We use our formalisation to translate $A$ into an SMT formula $\sem{A}_f$; we translate opcodes sequentially as per the ACIR program, conjoining the constraints generated for each opcode. For each verification condition in the ACIR program, these are \lstinline{BrilligCall} opcodes capturing \assert{$cond$} statements in the Noir program, we recover the boolean ACIR \lstinline{Expression} $exp_{cond}$ representing the $cond$, and we create the formula $F = \sem{A}_f \land \lnot \sem{exp_{cond}}_f$. The modular integer encoding can be used instead of the finite field one, if preferred, of course. We then use the cvc5 SMT solver to check the satisfiability of each of these $F$ formulas. Unsatisfiability means that the verification condition holds. A satisfiable formula, however, represents the failure of the corresponding verification condition, namely, there is a program execution that falsifies the corresponding predicate at the given program point. In this case, cvc5 generates a model, i.e. a satisfying assignment to the witnesses (including the ones representing the input parameters of the Noir program). This model represents a counterexample detailing how the verification condition is falsified, which is passed back to the developer to help debug the program. For instance, in Figure~\ref{fig:basic_program}, we use \nave{} to verify that the program input \lstinline{x} is either $0$ or $1$; it checks that the \lstinline{x * (x - 1) == 0} constraining works as expected, and \lstinline{a | b} captures a disjunction between \lstinline{a} and \lstinline{b}. If one of the disjuncts in the verification condition is removed, however, \nave{} returns a model demonstrating that \lstinline{x} can take that value too, namely, if we remove \lstinline{(x == 1)}, \nave{} offers a counterexample where \lstinline{x} $= 1$.

\subsection{Evaluation}

We evaluate our Noir program formal verifier in practice. We analyse its performance on some test programs that are part of the Noir language infrastructure. They come from the Noir repository, but we have annotated them with verification conditions. We tried to collect programs that had a mix of complexity and adhered to the ACIR subset that we have formalised. We have created sets of test programs with both verified and falsified conditions to analyse the behaviour of the solver with these two distinct types of searches. Moreover, we have also created sets of programs with and without range checks  --- i.e. \lstinline{BlackBoxFuncCall} opcodes calling function $Range(x,n)$. We wanted to analyse the performance of the solver, and consequently of our verifier, in analysing the type of complex bitsum constraints they induce. We have carried out our evaluation on a dedicated machine with Intel Core i7-9750H CPU @ 2.60GHz x 12, and 16GB of RAM, running Ubuntu 22.04.5 LTS. The test programs that we used and instructions on how to reproduce our evaluation are available in our repository.\footnote{\url{https://github.com/pedrotbtl/nave}}


\begin{table}[!b]
\centering
\begin{minipage}{0.4\textwidth}
\begin{tabularx}{\linewidth}{|X|c|c|c|c|c|}
\hline
\multicolumn{6}{|c|}{\textbf{Verified}} \\
\hline
\multicolumn{3}{|c|}{\textbf{Benchmark}} & \multicolumn{3}{c|}{\textbf{Time}(s)} \\
\hline
\textbf{id} & \textbf{\#w} & \textbf{\#o} & \textbf{sp} & \textbf{gb} & \textbf{int} \\ \hline
$pnv_1$ & 1 & 1 & 17.5 & 20.3 & 17.8 \\ \hline
$pnv_2$ & 1 & 2 & 16.0	& 22.3 & 17.3 \\ \hline
$pnv_3$ & 6 & 6 & 18.9	& 20.2 & $t$ \\ \hline
$pnv_4$	& 17	& 18 & 16.2	& 18.8	& t \\ \hline
$pnv_5$ & 1 & 1 & 17.5	& 17.1 & 19.0 \\ \hline
$pnv_6$ & 207 & 139 & 18.5	& 17.9 & 17.2  \\ \hline
$pnv_7$ & 1 & 1 & 18.2	& 16.8 & 20.3 \\ \hline
$pnv_8$ & 285 &	286 & 19.2 &	17.1 & 19.7 \\ \hline
$pnv_9$ & 4 &	4 &	18.4	& 16.7 & $t$ \\ \hline
$pnv_{10}$ &	1 &	1 &	19.0	& 15.7 & 17.8 \\ \hline
$pnv_{11}$ & 1 & 2 & 20.1	& 17.0 & 17.5 \\ \hline
$pnv_{12}$ &	5 &	4 &	17.4	& 17.4	& 15.2 \\ \hline
\end{tabularx}
\end{minipage} \hspace{.5cm}
\begin{minipage}{0.4\textwidth}
\begin{tabularx}{\linewidth}{|X|c|c|c|c|c|} \hline
\multicolumn{6}{|c|}{\textbf{Falsified}} \\
\hline
\multicolumn{3}{|c|}{\textbf{Benchmark}} & \multicolumn{3}{c|}{\textbf{Time}(s)} \\
\hline
\textbf{id} & \textbf{\#w} & \textbf{\#o} & \textbf{sp} & \textbf{gb} & \textbf{int} \\ \hline
$\textit{pnf}_1$	& 189	& 126	& 42.5	& 22.9 & $t$  \\ \hline
$\textit{pnf}_2$	& 5	& 4	& 18.9	& 17.9 & $t$  \\ \hline
$\textit{pnf}_3$	& 3	& 2 & 16.0	& 18.2 & 16.6  \\ \hline
$\textit{pnf}_4$	& 6	& 4	& 17.8	& 18.0 & 19.1  \\ \hline
$\textit{pnf}_5$	& 9	& 10 & 17.3	& 19.1 & $t$  \\ \hline
$\textit{pnf}_6$	& 20 & 3	& 18.1	& 18.9 & 16.0 \\ \hline
$\textit{pnf}_7$	& 4	& 3	& 17.9	& 17.6  & 15.0 \\ \hline
$\textit{pnf}_8$	& 12 & 6 & 17.6	& 18.9 & 14.8 \\ \hline
$\textit{pnf}_9$	& 2	& 2	& 16.8 &	18.7 & 14.8 \\ \hline
$\textit{pnf}_{10}$	& 9	& 5 & 17.1 &	18.4 & 14.6 \\ \hline
$\textit{pnf}_{11}$	& 1	& 1	& 17.5	& 18.8 & 16.0 \\ \hline
$\textit{pnf}_{12}$	& 2	& 2 & 16.3	& 18.5 & $t$  \\ \hline
$\textit{pnf}_{13}$	& 63 & 64 & 16.6	& 16.5 & 17.4  \\ \hline
\end{tabularx}
\end{minipage}
\caption{Results for Noir programs without range constraints with verification conditions verified and falsified, respectively. The column \textbf{id} identifies the program being tested, column \textbf{\#w} gives the program's number of witnesses, and column \textbf{\#o} the number of ACIR opcodes excluding \lstinline{BrilligCall}s. Columns \textbf{sp} and \textbf{gb} give the program analysis time for our ACIR finite-field encoding, whereas \textbf{int} gives the time for our integer encoding; we use $t$ to represent a 120-second timeout. For column \textbf{sp}, we used cvc5 solver with configuration \lstinline{--ff=split}, while for column \textbf{gb}, we used \lstinline{--ff=gb}.}
\label{tab:no_range}
\end{table}

In Table~\ref{tab:no_range}, we present the results for running \nave{} in two sets of programs without range check constraints: one set has programs with verified verification conditions, and the other one falsified conditions. We test both of our encodings. For the finite-field one, we also tested two cvc5 finite-field solver configurations: \lstinline{--ff=split} and the default \lstinline{--ff=gb}. These two configurations deploy two different solvers to check for the satisfiability of a finite-field theory formula. Simply put, the technique used by the former (described in~\cite{Ozdemir24}) approximates the one used by the latter (described in~\cite{Ozdemir23}); these approximations can be very effective in tackling certain combinations of bitsum constraints. The results demonstrate that there is no clear winner between our three \nave{} backends: encoding with integer theory, and finite-field theory with both cvc5 configurations. For example, there are test programs for which the integer encoding is the quickest, and there are others for which it is the sole backend to timeout. The two finite-field solvers behave very similarly; this is expected, given that the \lstinline{--ff=split} solver is meant to improve on bitsum constraints, and these programs have no range checks, i.e. the opcode leading to these types of constraints.

\begin{table}[!b]
\centering
\begin{minipage}{0.48\textwidth}
\begin{tabularx}{\linewidth}{|X|c|c|c|c|c|c|c|}
\hline
\multicolumn{8}{|c|}{\textbf{Verified}} \\
\hline
\multicolumn{5}{|c|}{\textbf{Benchmark}} & \multicolumn{3}{c|}{\textbf{Time}(s)} \\
\hline
\textbf{id} & \textbf{\#w} & \textbf{\#o} & \textbf{\#r} & \textbf{r$_{max}$} & \textbf{sp} & \textbf{gb} & \textbf{int} \\ \hline
$prv_1$	& 12	& 21	& 11 & 32	& 17.2	& 16.1 & 18.9 \\ \hline
$prv_2$	& 147	& 154	& 52 & 32	& 17.6	& 15.5 & 17.7  \\ \hline
$prv_3$	& 15	& 13	& 11 & 8	& 16.9	& 15.6 & 19.6  \\ \hline
$prv_4$	& 1	& 3 & 2	& 1 & 16.5	& 14.5 & 18.2 \\ \hline
$prv_5$	& 5	& 6 & 3	& 32 & 16.4	& 14.6 & 18.6  \\ \hline
$prv_6$	& 20	& 28	& 13 & 8 & 20.1	& 14.7 & 18.9  \\ \hline
$prv_7$	& 10	& 14	& 5	& 32 & 17.8	& 14.8 &  18.7 \\ \hline
$prv_8$	& 50	& 61	& 15 & 128 & 16.6	& 14.5 & 19.0  \\ \hline
$prv_9$	& 10	& 13	& 1	& 32 & 18.1	& 14.9 & $t$ \\ \hline
$prv_{10}$	& 969	& 984	& 292 & 8 & 17.6	& 15.9	& 19.9  \\ \hline
$prv_{11}$	& 74	& 86	& 26	& 238 & 19.2	& 15.7 & 18.8 \\ \hline
$prv_{12}$ & 21 & 37	& 9 & 32 & 18.0 & $t$ & $t$ \\ \hline
$prv_{13}$	& 8	& 6 & 1	& 1 & 16.4	&	18.5 & $t$  \\ \hline
$prv_{14}$	& 20	& 31 & 13	& 1 & 17.5	&	18.9	&	15.4 \\ \hline
\end{tabularx}
\end{minipage} \hfill
\begin{minipage}{0.47\textwidth}
\begin{tabularx}{\linewidth}{|X|c|c|c|c|c|c|c|}
\hline
\multicolumn{8}{|c|}{\textbf{Falsified}} \\
\hline
\multicolumn{5}{|c|}{\textbf{Benchmark}} & \multicolumn{3}{c|}{\textbf{Time}(s)} \\
\hline
\textbf{id} & \textbf{\#w} & \textbf{\#o} & \textbf{\#r} & \textbf{r$_{max}$} & \textbf{sp} & \textbf{gb} & \textbf{int} \\ \hline
$\textit{prf}_1$	& 24	& 31	& 15 & 1	& 18.3	&	18.0	&	16.3 \\ \hline
$\textit{prf}_2$	& 3	& 4	& 3	& 32 & 18.8	&	19.3	&	16.8 \\ \hline
$\textit{prf}_3$	& 4	& 4	& 2	& 32 & 17.2	&	18.5 & $t$ \\ \hline
$\textit{prf}_4$	& 7	& 6 & 1	& 1 & 17.5	&	17.8	&	17.4  \\ \hline
$\textit{prf}_5$	& 3	& 3	& 3	&  1 & 17.7	&	16.7	&	17.0 \\ \hline
$\textit{prf}_6$	& 5	& 7	& 1	& 1 & 18.0	&	17.6	&	16.0   \\ \hline
$\textit{prf}_7$	& 2	& 3	& 1	& 8 & 18.0	&	86.2	&	16.3   \\ \hline
$\textit{prf}_{8}$	& 3	& 4	& 1	& 1 & 16.5	&	17.9	&	16.9 \\ \hline
$\textit{prf}_{9}$ & 32	& 28 & 6	& 32 & 22.6	&	19.4	&	17.0  \\ \hline
$\textit{prf}_{10}$ &	9	& 16 &	7 & 32 &	57.9 & $t$ & $t$ \\ \hline
$\textit{prf}_{11}$ &	6	& 13 &	7	& 32 & 28.6 & $t$ & $t$ \\ \hline
\end{tabularx}
\end{minipage}
\caption{Results for Noir programs without range constraints with verification conditions verified and falsified, respectively. The column \textbf{id} identifies the program being tested, column \textbf{\#w} gives the program's number of witnesses, column \textbf{\#o} the number of ACIR opcodes excluding \lstinline{BrilligCall}s, \textbf{\#r} the number of \lstinline{BlackBoxFuncCall} range-check opcodes, and \textbf{r$_{max}$} the number of bits for the largest range checked. Columns \textbf{sp} and \textbf{gb} give the program analysis time for our ACIR finite-field encoding, whereas \textbf{int} gives the time for our integer encoding; we use $t$ to represent a 120-second timeout. For column \textbf{sp}, we used cvc5 solver with configuration \lstinline{--ff=split}, while for column \textbf{gb}, we used \lstinline{--ff=gb}.}
\label{tab:range}
\end{table}

In Table~\ref{tab:range}, we present the results for two sets of programs \emph{with} range check constraints. Again, one set has programs with verified verification conditions, and the other one has falsified conditions, and we test both encoding and solver configurations for the finite-field encoding. Again, the results demonstrate that there is no clear winner between our three \nave{} backends. However, we do see a pattern: range checks for larger ranges --- i.e. for representations with larger number of bits --- take longer, timing out or taking significantly longer than other test programs. Also, we can see a difference in the behaviour of the finite-field solvers: the \lstinline{--ff=split} solver significantly outperforms the \lstinline{--ff=gb} solver in some instances (see \textit{prv$_{12}$}, \textit{prf$_7$}, \textit{prf$_{10}$}, and \textit{prf$_{11}$}).

We illustrate two of the test programs used in our evaluation in Figure~\ref{fig:test_programs}. They are both regression tests for the Noir language. The left-hand-side program is an excerpt of the \textit{prv$_{13}$} --- we removed some code for the sake of conciseness. The \lstinline|verify_assert| statement that we added to this program captures restrictions on the valid values of \lstinline|x| and \lstinline|y| based on the two \lstinline|assert|s in the code; \lstinline{a & b} captures a conjunction between \lstinline{a} and \lstinline{b}. The first two conjuncts in the \lstinline|verify_assert| are enforced by the first \lstinline|assert|, and the following two are enforced by the second \lstinline|assert|. The right-hand-side program (\textit{prv$_{14}$})  relies on range constraints imposed by the 1-bit unsigned integer \lstinline|u1| type. The \lstinline|verify_assert| statement we added to this program captures some restrictions that are enforced by the assignments to variables \lstinline|u|, \lstinline|v|, and \lstinline|p|.

\begin{figure}[!t]
    \centering
    \begin{minipage}[t]{0.45\textwidth}
\begin{lstlisting}
fn main(x: Field) {
    let y = unsafe { 
        empty(x + 1) };
    let z = y + x + 1;
    let z1 = z + y;
    assert(z + z1 != 3);
    let w = y + 2 * x + 3;
    assert(w + z1 != z);
    unsafe { 
    verify_assert(
        ((y != 1/3) | (x != 0)) 
        & ((y != 0) | (x != 1/2)) 
        & ((y != -1) | (x != -1/2)) 
        & ((y != -1/2) | (x != -1))
    );}
}

unconstrained fn empty(_: Field) 
    -> Field { 0 }
\end{lstlisting}    
\end{minipage}\hfill
\begin{minipage}[t]{0.45\textwidth}
\begin{lstlisting}
fn main(x: u1, y: u1, z: u1) 
    -> pub u1 {
    let p = y - z;
    if p != 0 { 
        let a = x / z;
        let b = a - z;
        if b != 0 {
            let _ = a / b;
        }
    }
    let u = x - z;
    let v = y - u;
    unsafe { 
    verify_assert(
        ((u != 0) | (x == z)) 
        & ((v != 0) | (u == y))
        & ((p != 0) | (y == x))
    );}
    v
}
\end{lstlisting}
\end{minipage}
    \caption{Regression test programs without (excerpt of $\textit{prv}_{13}$ with range check removed) and with ($\textit{prv}_{14}$) range-check opcode, respectively.}
    \label{fig:test_programs}
\end{figure}

Our results suggest that we could use our encodings and the different solvers for finite-field encodings in a complementary way. We could use the three backends concurrently for the same program, reporting the results of whichever one finishes first. Furthermore, the results suggest that programs with large range checks can take a significant amount of time to be verified. This demonstrates that abstractions (i.e. approximations) need to be investigated to address this issue. These results chart a path for future work where we will investigate this concurrent combination of solvers and abstractions to tackle complex constraints.




\section{Related Work}
\label{sec:related}
In this section, we discuss the literature related to our verification framework.

\textit{ZK programming languages and compilers.} CirC~\cite{Ozdemir22} provides an infrastructure for compiling programs into circuits. The program is first converted to an intermediate representation called CirC-IR. This IR is an abstract existentially quantified circuit (EQC) based on the SMT-LIB standard and supports booleans, numbers, bit vectors, and arrays. Thus, CirC is somewhat similar to LLVM but is targeted at ZKP DSLs. The infrastructure can support compilation of cryptographic languages to CirC-IR; however, integrating a new language requires implementing a corresponding parser and interpreter. CirC has been evaluated on Circom~\cite{circom-lang}: this is arguably the most popular (low-level) arithmetic circuit DSL. The main difference between NAVe and CirC is that NAVe operates on ACIR, which already resembles CirC-IR, eliminating the need for conversion. Also, ACIR support some complex higher-level constructs, like black box functions and memory manipulations, underpinning Noir, which Circ-IR does not.

Like Noir, Cairo~\cite{cairo-lang} is another popular high-level ZK programming language. So far, there has been very little work on the formal verification of Cairo programs. The only paper in this area that we found~\cite{Avigad22} formalises and proves the soundness of a transformation of Cairo programs into an algebraic representation used by a ZK proof system. Our verifier and formalisation seem to be the first to deal with a high-level ZK programming language and its particularities. For instance, in Noir, we have complex constructs in the form of black box functions and a notion of memory manipulation.

\textit{Static analysis and bug detection.} 
The majority of the existing work here has focused on Circom and other circuit DSLs. Some claim that their approaches can generalise beyond Circom, but their evaluations are limited to Circom circuits. 
Noir also compiles its Rust-style DSL to ACIR, making it prover-agnostic.
Z{\scalebox{0.7}{EQUAL}}~\cite{Stephens25} is a framework for verifying ZK circuits using static analysis and deductive verification. It is built on top of the Circom compiler and performs SMT queries via Z3 using a modular integer arithmetic encoding of finite-field constraints. It can analyse circuit templates that do not require full instantiation. However, the key difference from NAVe is that Z{\scalebox{0.7}{EQUAL}} also operates at the DSL level and is evaluated only on Circom circuits.
ZK{\scalebox{0.7}{AP}}~\cite{Wen24} is also a static analysis framework to detect vulnerabilities in ZK circuits. ZK{\scalebox{0.7}{AP}}'s approach is guided by a manual study of existing bugs in Circom programs. It uses a circuit dependence graph to capture key circuit properties and expresses semantic vulnerability patterns as Datalog-like queries. They have implemented nine different detectors, each identifying a specific anti-pattern. QED\textsuperscript{2}~\cite{Pailoor23} also uses SMT solvers for automated verification, focusing on particular circuit properties (underconstrained bug to be precise) rather than full functional correctness. Although QED\textsuperscript{2} can, in principle, analyse polynomial equations over finite fields, it has been evaluated only on Circom circuits.

\textit{Verification of Noir programs.} Lampe \cite{lampe} is a project that aims to support the formal verification of both the semantics of the Noir language and the properties of programs written in Noir. It models Noir's semantics in the Lean \cite{Lean4} programming language and theorem prover, focusing more on capturing the semantics of Noir rather than on providing a verification framework. Given its use of a theorem prover, Lampe users have to iteratively drive the proving of properties, whereas \nave{} uses SMT solving to automatically check them.

\section{Conclusion}
\label{sec:conclusion}

Motivated by the recent increase in popularity of ZK programs and the emergence of high-level ZK programming languages, we propose \nave{}: a framework for the formal verification of Noir programs. We propose a formalisation for Noir via ACIR, an intermediate representation for Noir programs. We propose two SMT-LIB-based formalisations of ACIR programs: one using a non-standard finite-field theory, and another using the standard theory of integers. Our formalisation tackles some peculiarities of the Noir/ACIR language that are not present in other languages. For instance, ACIR has several black box functions with complex behaviour. We started tackling the range-check black box function, but we intend to expand our encoding to other ones. The memory operations of ACIR are another feature that is not present in other traditional languages. Moreover, we have to address particularities of the ZK paradigm, such as the distinction between constrained and unconstrained code. 

We use our formalisation to create the \nave{} formal verifier. It is implemented as the \lstinline{formal-verify} command within a fork of Nargo, Noir's popular package manager. We hope that this integration motivates Noir developers to use our tool. We rely on the Noir compiler to generate the ACIR representation of a program, which is later SMT-LIB encoded using our formalisation and checked for satisfiability by cvc5. The main goal of our tool is to identify underconstrained programs, a common affliction of ZK programs. These are programs that have some constraints missing, possibly due to the unrestricted use of unconstrained code. \nave{} users can specify verification conditions using the \assert{$\cdot$} function. Our tool also reports counterexamples to developers for verification conditions that are falsified, helping them debug their code in the process. We have evaluated \nave{} on some test programs that are part of the Noir language infrastructure with promising results. They also demonstrate that, in some cases, better abstractions are needed to tackle complex constraints, such as the bitsum ones generated by range checks.

As future work, we plan to extend our formal verifier to tackle a larger subset of ACIR. We will particularly investigate how we can formalise other types of black box functions; these are notoriously difficult to encode. Moreover, we plan to investigate new abstractions to tackle complex ACIR constraints/programs.


\bibliographystyle{plain}
\bibliography{references}

\end{document}